\documentclass[a4paper]{spie}  
\usepackage[dvips]{graphicx}

\title{SPOrt: an Experiment Aimed at Measuring the Large Scale Cosmic 
Microwave Background Polarization} 

\author{Ettore~Carretti,\supit{a}
Stefano~Cortiglioni,\supit{a}
Gianni~Bernardi,\supit{a}
Stefano~Cecchini,\supit{a}    \and
Claudio~Macculi,\supit{a}
Carla~Sbarra,\supit{a}
Jader~Monari,\supit{b}
Alessandro~Orfei,\supit{b}
Marco~Poloni,\supit{b}    \and
Sergio~Poppi,\supit{b}
Giuliano~Boella,\supit{c}
Silvio~Bonometto,\supit{c}
Massimo~Gervasi,\supit{c}
Giorgio~Sironi,\supit{c}    \and
Mario~Zannoni,\supit{d}
Marco~Tucci,\supit{e}
Massimo~Baralis,\supit{f}
Oscar~A.~Peverini,\supit{f}
Riccardo~Tascone,\supit{f}    \and
Giuseppe~Virone,\supit{f}
Roberto~Fabbri,\supit{g}
Luciano~Nicastro,\supit{h}
Kin-Wang~Ng,\supit{i}
V.A.~Razin,\supit{j}    \and
Evgenij~N.~Vinyajkin,\supit{j}
Mikhail~V.~Sazhin,\supit{k}
Igor~A.~Strukov\supit{l}    \and
\skiplinehalf
\supit{a}IASF/CNR Sezione di Bologna, Via Gobetti 101, I-40129 Bologna, Italy \\
\supit{b}IRA/CNR, Via Gobetti 101, I-40129, Bologna, Italy\\
\supit{c}Dip. di Fisica, Univ. di Milano - Bicocca, 
     P.za della Scienza 3, I--20126 Milano, Italy\\
\supit{d}IASF/CNR Sezione di Milano, Via Bassini 15, I-20133 Milano, Italy \\
\supit{e}Instituto de Fisica de Cantabria, Avda Los Castros s/n,
     39005 Santander, Spain\\
\supit{f}IRITI/CNR, c.so Duca degli Abruzzi 24, I--10129 Torino,
     Italy\\
\supit{g}Dip. di Fisica, Univ. di Firenze, Via Sansone 1, 
           I--50019 Sesto Fiorentino (FI), Italy\\
\supit{h}I.A.S.F./C.N.R. Sez. di Palermo, via U. La Malfa 153, 
     I--90146 Palermo, Italy\\
\supit{i}Academia Sinica, 11529 Taipei, Taiwan\\
\supit{j}NIRFI, 25 B.Pecherskaya st., Nizhnij Novgorod 603600/GSP-51,
      Russia\\
\supit{k} Shternberg Astronomical Institute, 
           Moscow State University, 
           Moscow 119992, Russia\\
\supit{l}IKI, Profsojuznaja ul. 84/32, Moscow 117810, Russia\\
}

\authorinfo{Further author information: (Send correspondence to E.C.)\\
E.C.: E-mail: carretti@bo.iasf.cnr.it, Telephone: +39 051 6398735}

\begin{document} 
  \maketitle 

\begin{abstract}
SPOrt (Sky Polarization Observatory) is a space experiment to be flown on the
International Space Station during Early Utilization Phase aimed
at measuring the microwave polarized emission with FWHM=7$^\circ$, 
in the frequency range 22-90~GHz.
The Galactic polarized emission can be observed at the lower frequencies 
and the polarization of Cosmic Microwave Background (CMB) 
at 90 GHz, where contaminants are expected to be less important.
The extremely low level of the CMB Polarization signal ($< 1$~$\mu$K) 
calls for intrinsically stable radiometers.
The SPOrt instrument is expressly devoted to CMB polarization measurements
and the whole design has been optimized for minimizing instrumental
polarization effects.
In this contribution we present the receiver architecture based on
correlation techniques, the analysis showing its intrinsic stability and
the custom hardware development carried out to detect such a low signal.
\end{abstract}



\section{Introduction}\label{intro}
The Cosmic Microwave Background (CMB) is a powerful tool for
understanding the origin and evolution of the Universe. The hot Big Bang model
predicts the CMB is a Black Body radiation almost isotropic and unpolarized.
Any deviations from this ideal behaviour is related to
cosmological
parameters and allow their 
determination\cite{sazhin95,jungman96,zalda97,efstathiou99}. 
Very small CMB Anisotropies (CMBA) has been detected
at both large\cite{sm91,be96} and small\cite{DeBe00,Ha00,MiCa99}
angular scales, but only upper limits on the CMB Polarization (CMBP)
have been established up to now.
\begin{figure} [bht]
\includegraphics[width=0.5\linewidth, angle=0]{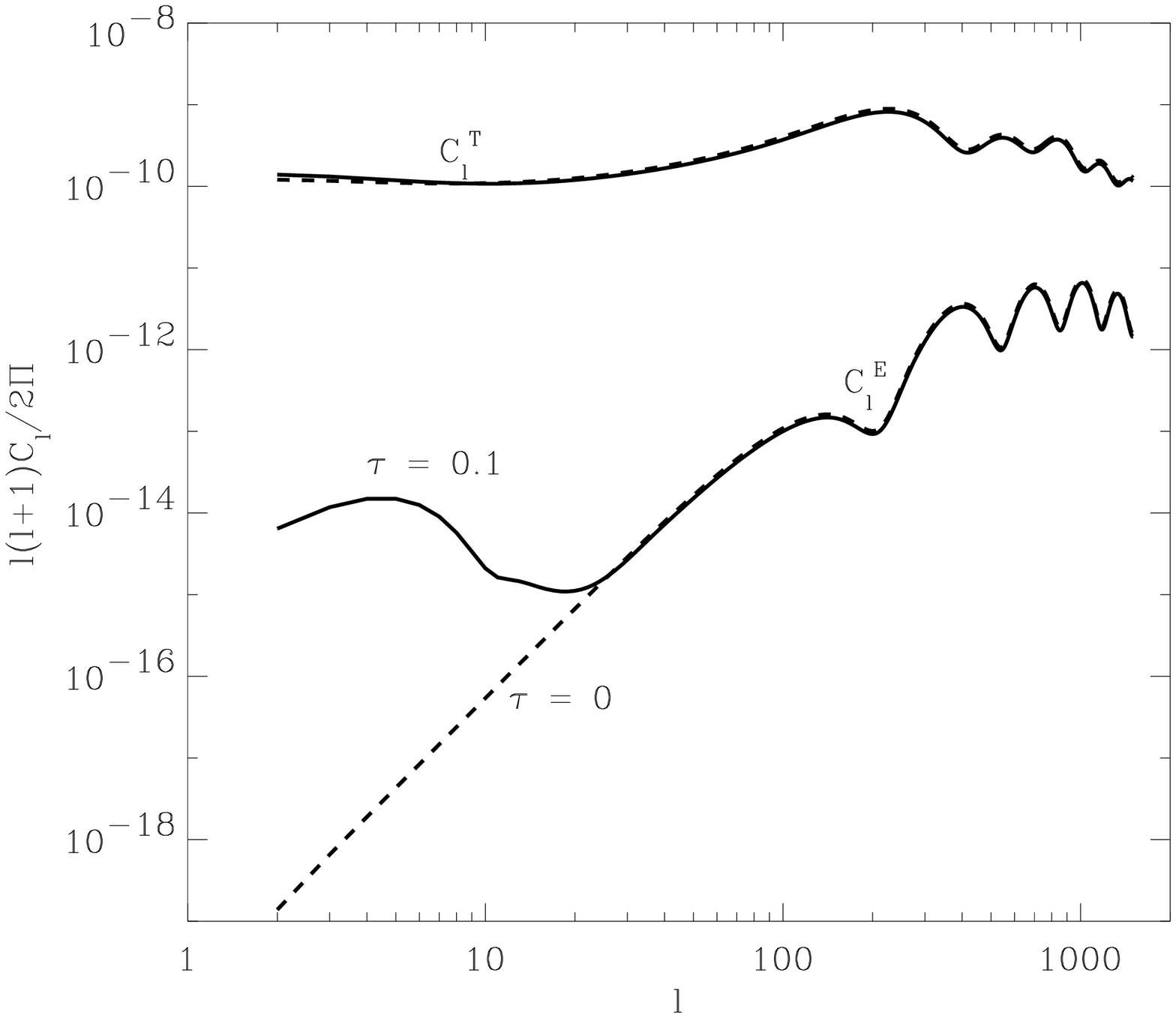}
\includegraphics[width=0.5\linewidth, angle=0]{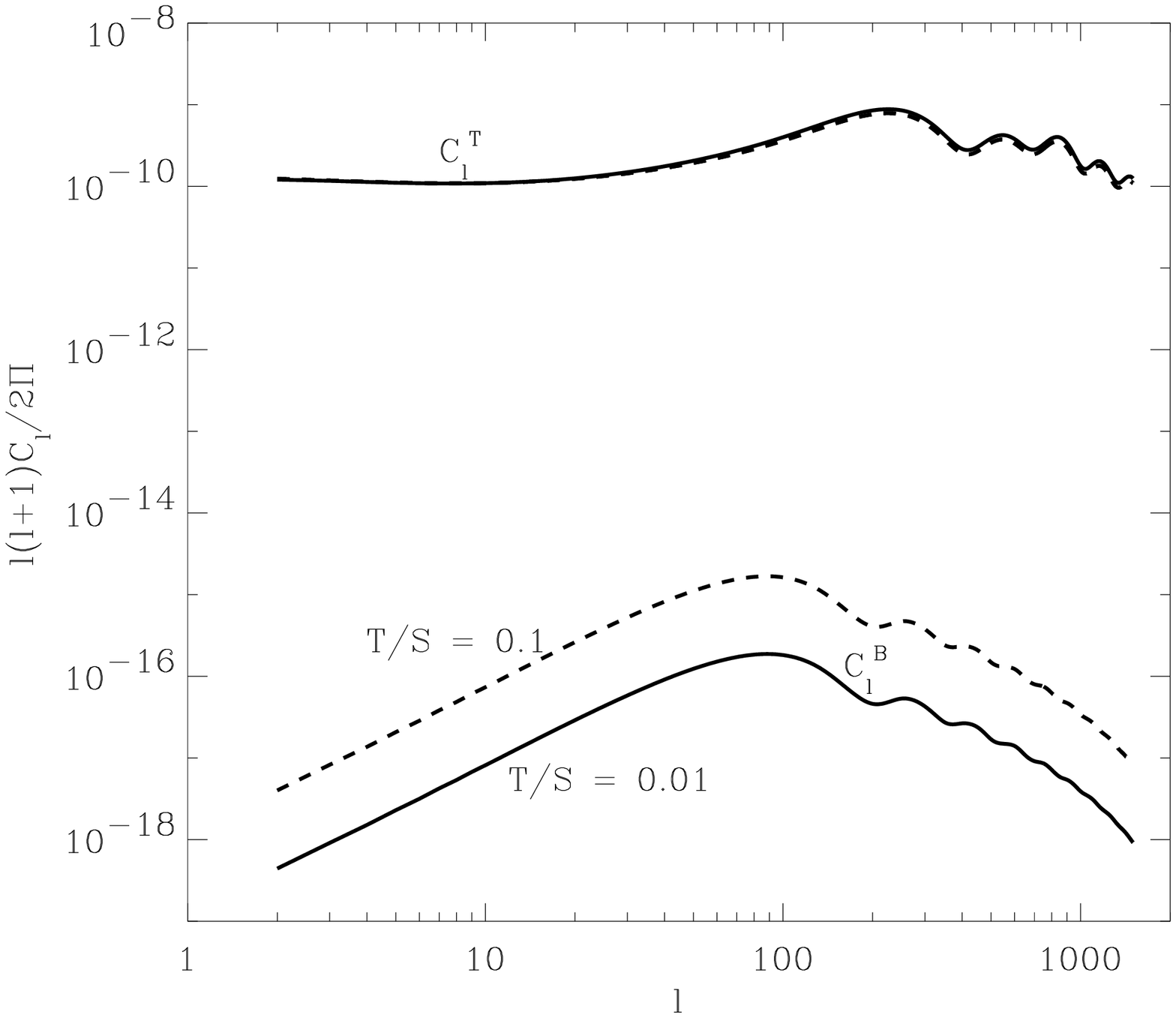}
\caption{{\bf Left:} Temperature anisotropy and E-mode power spectra. 
Two $\Lambda$CDM ($\Omega_{\Lambda}=0.7$) cosmological
models which differ only in the re-ionization 
optical depth $\tau$ are shown. {\bf Right:}
Temperature anisotropy and B-mode power spectra. 
Two $\Lambda$CDM ($\Omega_{\Lambda}=0.7$) cosmological
models which differ only in the tensor (gravitational waves) 
to scalar perturbation ratio T/S  
are shown.}
\label{CTCECBfig}
\end{figure}
In spite of its elusiveness, the CMBP promises to add information
to the CMBA data: it will provide a direct measurement of 
cosmological parameters that CMBA 
alone is not able to determine\cite{zalda97}. 
The optical depth $\tau$ in the dark ages and the epoch $z_{\rm ri}$ at which
the re-ionization occurred are directly measured by CMBP providing us with
the formation epoch of the first structures and their growth-rate.
Figure~\ref{CTCECBfig} presents a 
comparison between temperature and
polarization (E-mode) power spectra for two cosmological models, which differ
only in the optical depth $\tau$ of the re-ionized medium. 
It is clear that the E-mode spectrum is much more 
sensitive to $\tau$ than the temperature spectrum and that this 
new information is found at
large angular scales ($l<10$, i.e. $\theta > 20^{\rm o}$). 
In addition, the E-mode of the CMBP brings important information also 
at subdegree angular scales, where
the {\it coherent} primordial fluctuations
predicted by inflation leave fingerprints like a well defined Doppler peak 
pattern: the peaks in the T spectrum correspond to minima in the E spectrum
and viceversa. Thus, the detection of the CMBP at subdegree scales leads 
to an indirect check of the inflationary model\cite{Ko99}.

The detection of the B-mode is even more exciting, although the 
signal is very weak: 
its level
is directly related to the tensor-to-scalar perturbation ratio T/S 
(see Figure~\ref{CTCECBfig}), 
whose value is in turn
related to the energy of the Universe at the inflation time\cite{Ru82,KaKo98}. 
Thus, the measurement of the B-mode allows the estimate of the energy at 
which the inflation occurred and the identification of the right model in
 the zoo of the existing ones.

Moreover, besides their intrinsic interest, $\tau$, $z_{\rm ri}$ and T/S
determinations further improve the precision on other cosmological parameters.

Unfortunately, the CMBP predicted level is very low 
(few $\mu$K on sub-degree scales and less than 1~$\mu$K on large scales).
Current experimental upper limits are still one order of magnitude higher than
the predicted 
level\cite{penzias65,caderni78,nanos79,lubin81,partridge88,wollack93,netterfield95,sironi98,subrahmanyan00,keating01,hedman02}
and, even though more sensitive detectors are coming, CMBP measurements are
biased by foreground subtraction.
Besides its intrinsic interest, the Galaxy 
acts as a foreground for CMB experiments and only its accurate 
knowledge will allow measurements of CMB features 
(See Ref.~\citenum{Tu00,Teg00,Br02} and references therein).
So far, polarization surveys have been carried out only
at frequencies up to 2.7 GHz\cite{BrSp76,Du97,Du99,Uy99,Ga01}, where
the Galactic emission appeares to be dominated by synchrotron. 
Such observations either
are widely undersampled\cite{BrSp76} or cover narrow 
stripes around the Galactic Plane\cite{Du97,Du99,Uy99,Ga01}.
A better
estimate of the foreground contaminations can be done only through 
Galactic surveys at frequencies closer to the range of interest of 
CMBP measurements.

\section{The SPOrt Experiment}

The purpose of Sky Polarization Observatory 
(SPOrt){\footnote{http://sport.bo.iasf.cnr.it}}
is aimed at filling
the current gap in measurements of the 
diffuse polarized emission in the 22-90~GHz range.
Together with BaR-SPOrt, on-ground observations and
technological activities, it is part of the SPOrt Programme\cite{Co02}
aimed at detecting the CMBP.
SPOrt is an Italian Space Agency (ASI) 
funded experiment and it has been selected by ESA to be flown 
onboard the International 
Space Station (ISS) for a minimum lifetime
of 18 months, starting from 2005 (see Figure~\ref{columbusfig}).
\begin{figure} [bht]
\vskip 6cm 
\caption{SPOrt position on the External Payload Facility of Columbus 
onboard the ISS (courtesy by Alenia Spazio).}
\label{columbusfig}
\end{figure}

SPOrt is the first space instrument devoted to $Q$~\&~$U$ Stokes 
parameters measurements in the microwave domain at large angular scales
($\theta > 7^{\rm o}$).
This can be done only by all-sky surveys (space experiments) and by instruments
designed  to be as much as possible insensitive to instrumental polarization.  

The main features of SPOrt are the following  (see also Table~\ref{tabfeat}):
\begin{itemize}
   \item multifrequency approach with four frequency channels 
         at 22, 32, 60 and 90~GHz to match the 
         best band for CMBP observation (90~GHz), while checking the Galactic 
	 contributions (22-90~GHz) and mapping the Galactic 
	 synchrotron emission (22 and 32~GHz).
   \item very simple optics (corrugated feed horns), providing angular
         resolutions down to $7^{\circ}$, suitable to access the new 
         information on
	 cosmological parameters contained in CMBP on large scales, while
	 minimizing optics systematic effects.
   \item a nearly all-sky survey ($\sim80$\% sky coverage).
\end{itemize}
\begin{table}[h]
\caption{SPOrt main features: $N_{PX}$ is the number of FWHM 
pixels covered by SPOrt,
$\sigma_{1s}$ 
is the istantaneous sensitivity (1 second),  
$\sigma_{PX}$ and $\sigma(P_{\rm rms})$ 
are the final sensitivity per pixel and for the 
$P_{\rm rms} = \sqrt{\left<Q^2 + U^2\right>}$,  
respectively, considering a 18~month mission lifetime and 50\% observing
efficiency.} 
\label{tabfeat}
\begin{center}       
\begin{tabular}{ccccccccc} 
\hline
\rule[-1ex]{0pt}{3.5ex}
  {$\nu$ [GHz]} &
  {BW} & 
  {FWHM} &
  {Orbit Time [s]}&
  {Coverage}&
  {$N_{PX}$}&
  {$\sigma_{1s}[{\rm mK}{\rm s}^{1/2}]$} &
  {$\sigma_{PX}[{\rm\mu K}]$} &
  {$\sigma(P_{\rm rms})[{\rm\mu K}]$} \\
\hline
22, 32, 60, 90 & 10\% & $7^\circ$	&  5400 & 
80\% & 660 & 1.0 & $5.2$ & $0.3$ \\ 
\hline
\end{tabular}
\end{center}
\end{table}


The CMBP signal is weak (about 1-10\% of CMBA, 
depending on the scale) requiring expressly devoted 
instruments, as CMBA does. Thus, great care has been taken
to optimize the instrument design with respect to systematics generation,
long term stability and observing time efficiency. 

The following major choices
were adopted for the SPOrt design:
\begin{enumerate}
   \item correlation polarimeters to improve the stability 
         (see Figure~\ref{radioFig});
   \item correlation of the two circularly polarized components $E_L$ and $E_R$ 
         to directly and simultaneously measure 
	 both $Q$ and $U$ (100\% observing time efficiency)
         \begin{eqnarray}
             Q &\propto& \Re (E_R E_L^*) \nonumber\\
             U &\propto& \Im (E_R E_L^*) 
              \label{qulreq}
         \end{eqnarray}
         This optimizes
         the sensitivity with respect to other schemes 
         which provide either $Q$ or $U$ at once, just resulting in 50\% 
         efficiency (e.g. correlation or difference of the
         linear components);
   \item detailed analysis of the correlation scheme to minimize the
         instrumental systematics by the identification of
         "critical" components and the specifications they have to satisfy;
   \item custom development of these components when the existing 
         state-of-the-art is not enough.
\end{enumerate}

\section{Design Analysis}

The radiometer equation\cite{Wo95,Wo98} helps us find the parameters to be
controlled for minimizing systematic effects. In fact, in the
expression{\footnote {$T_{\rm sys}$, $T_{\rm offset}$ and $\Delta T_{\rm offset}$ are the 
system
temperature, the offset equivalent temperature and its fluctuation,
respectively; $G$ is the radiometer
gain, $\tau$ the integration time, $\Delta \nu$ the radiofrequency bandwidth
and $k$ a constant depending on the
radiometer type.}
\begin{equation}
  \Delta T_{\rm rms} = \sqrt{
                      {k^2 \,T_{\rm sys}^2 \over \Delta\nu\,\tau}+
               T_{\rm offset}^2
               \left({\Delta G \over G}\right)^2 +
               \Delta T_{\rm offset}^2
            }
\label{trmseq}
\end{equation}
the first term represents the white noise of an ideal and stable
radiometer, while
the second and the third terms are the gain and offset fluctuation 
effects, respectively, and represent the additional noise generated 
by instrument instabilities.
The ideal behaviour is preserved provided that the offset is kept 
under control.

Correlation receivers are intrinsically more stable because of their 
lower offset generation. Figure~\ref{radioFig} shows the scheme we 
adopted for the SPOrt radiometers. 
\begin{figure} [bht]
\begin{center}
\includegraphics[width=0.6667\linewidth, angle=0]{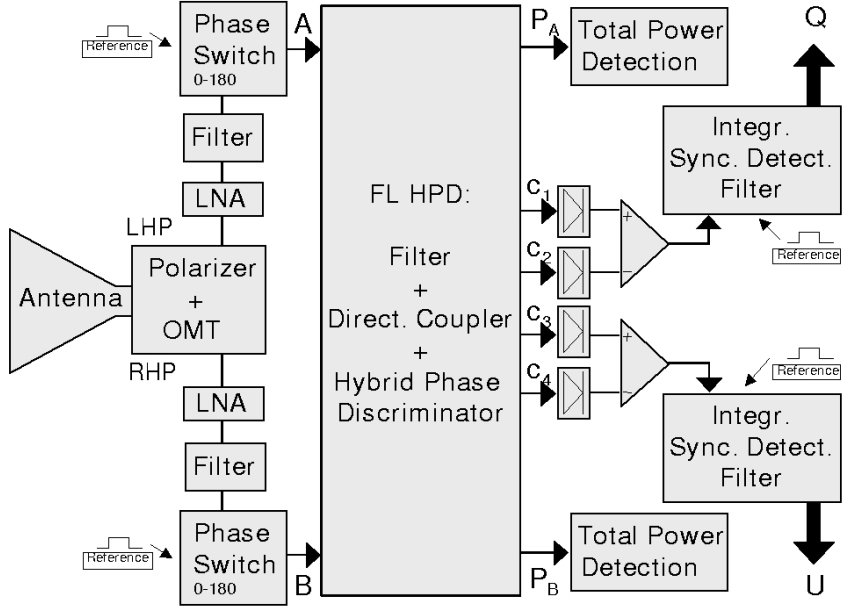}
\end{center}
\caption{Schematic block diagram of the SPOrt radiometers. Polarizer and OMT 
extract the two circularly polarized
components $LHP$~\&~$RHP$ collected by the horn. 
After amplification, the 
correlation unit (based on a Hybrid Phase Discriminator, 
see~\cite{Pe01,Pe02}
for details) provides directly both $Q$~\&~$U$.}
\label{radioFig}
\end{figure}
Polarizer and
OMT extract the two circularly polarized components collected by a 
dual-polarization feed horn. After amplification, the two components are
correlated by the correlation unit (CU). The latter includes an Hybrid Phase
Discriminator (HPD), diodes and differential amplifiers, whose 
outputs are the two Stokes parameters $Q$~\&~$U$.

In order to minimize the offset level an  
analysis has been carried out to identify the devices generating offset 
sources and the parameters to be controlled.
 The analysis shows the offset is generated at both CU
and antenna system (horn, polarizer and OMT) levels. 

The CU needs an HPD with high rejection of the unpolarized
components. This has been achieved by the development of a custom 
device\cite{Pe01,Pe02} 
providing $> 30$~dB rejection. In combination with a lock-in system, this makes
 the CU contribution to the offset negligible,  
the total rejection being $> 60$~dB.

Consequently, the antenna system is 
the most important offset source. 
Carretti et al. in Ref.~\citenum{Carr01} found that:{\footnote{
$T_{\rm sky}$ is the signal collected from the sky, 
$T_{\rm noise}^{\rm horn}$ is the noise generated by the horn only, 
  $T_{\rm noise}^{\rm Ant}$
is the noise temperature by the whole antenna system, $\eta$ is the
efficiency of the feed horn and $T_{\rm ph}^{\rm pol}$ is the physical
temperature of the polarizer.}}
\begin{eqnarray}
 T_{\rm offset}   =  S\!P_{\rm OMT}\left(T_{\rm sky} + 
                 T_{\rm noise}^{\rm Ant}
               \right) +
               S\!P_{\rm pol}
                  \left(T_{\rm sky} + 
                            T_{\rm noise}^{\rm horn} -
                            {T_{\rm ph}^{\rm pol} \over
                     \eta}
                          \right),
                  \label{AB0TNeq}
\end{eqnarray}
where the two quantities 
\begin{eqnarray}
 S\!P_{\rm OMT} & = & 2\,{\Re(S_{A1}S_{B1}^*)\over \left|S_{A1}\right|^2},
                     \\
 S\!P_{\rm pol} & = & {1\over 2} \left(1 - {\left|S_{\perp}\right|^2\over
                                 \left|S_{\parallel}\right|^2}\right),
                                 \label{sppoleq}
\end{eqnarray}
describe the goodness of the OMT and of 
the polarizer, respectively, from the offset generation
point of view. Uncorrelated signals (noise and sky) are partially
detected as correlated signals because of the OMT 
cross--talk ($S_{A1}$ and $S_{B1}$ are the transmission and cross-talk 
coefficient of the OMT, respectively) and of the polarizer
attenuation difference ($S_{\parallel}$ and $S_{\perp}$ are the attenuations
of the two polarization of the polarizer).

The instability of a radiometer can be measured in terms 
of the knee frequency ($f_{\rm knee}$), that provides the time 
scale at which the 1/$f$ component of the 
noise power spectrum prevails on the white noise.
Destriping techniques can remove most of the effects of the 1/$f$ noise, 
but only if the knee frequency is lower than
the signal modulation frequency\cite{Sbar02}. For SPOrt this corresponds to
the orbit frequency $f_{\rm orbit} = 1.8\times 10^{-4}$~Hz. 

Currently available InP Low Noise Amplifiers have rather high 
knee frequencies  ($f_{\rm knee}^{\rm lna}\sim$ 100-1000~Hz), 
making correlation architectures more convenient.
In fact, the knee frequency of a correlation receiver is related to that 
of its amplifiers by the formula 
\begin{equation}
f_{\rm knee} = \left({T_{\rm offset}\over T_{\rm sys}}\right)^2
                 f_{\rm knee}^{\rm lna} 
\end{equation}
where $T_{\rm offset}$ is the radiometric offset 
and $T_{\rm sys}$ is the system temperature.

Equations~(\ref{AB0TNeq})-(\ref{sppoleq}) state that the
main offset sources are the OMT cross-talk and 
the difference between the attenuations of the 
two polarizations of the polarizer.
SPOrt needs have been quantified in -60~dB and -30~dB, respectively, 
leading to an offset value
as low as $T_{\rm offset}\sim$~50~mK which, combined with a 
$T_{\rm sys} \sim$~100~K, gives the knee frequency
\begin{equation}
f_{\rm knee} \sim 2.5\times 10^{-7} f_{\rm knee}^{\rm lna} 
\end{equation}
matching the condition for a succesful destriping
($f_{\rm knee} < f_{\rm orbit}$)

However, state-of-the-art OMTs are not good enough, and custom 
hardware development has been required to the SPOrt team. 
Figure~\ref{crossfig} shows the result obtained for the 32~GHz channel:
a cross-talk as low as -65~dB has been achieved, well matching the -60~dB
goal.
\begin{figure} [htp]
\begin{center}
\includegraphics[width=0.6\linewidth, angle=0]{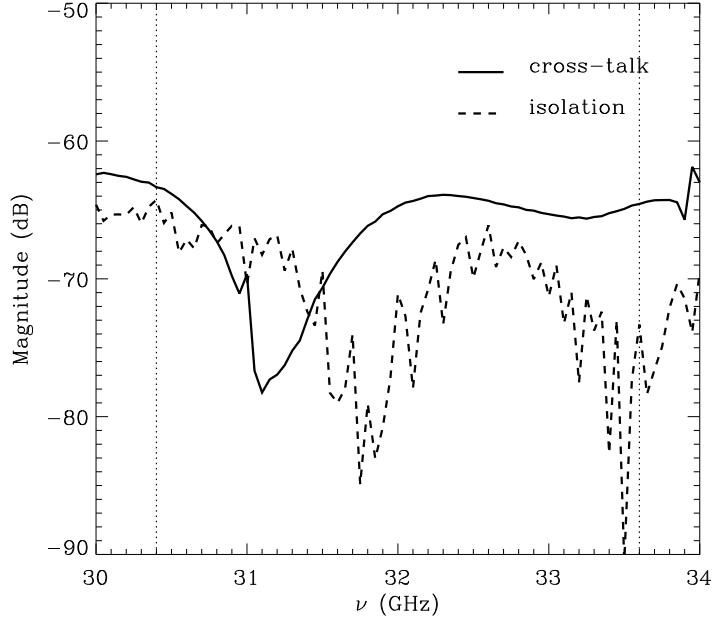}
\end{center}
\caption{Magnitudes of both the isolation between the two rectangular ports 
and the cross-talk between
the two polarizations for the 32~GHz SPOrt OMT. The vertical dotted lines 
show the 10\% band.}
\label{crossfig}
\end{figure}

Besides the offset generation, the SPOrt team has identified another 
systematic error source:
the spurious polarization generated by the optics\cite{Carr01}.
This is due to the 
anisotropy distribution of the unpolarized radiation 
modulated by the $f$ pattern:
\begin{eqnarray}
T^{\rm horn} &=&{1\over\Omega_A}\int_0^{\pi}\sin\theta\,d\theta
                                    \int_0^{\pi/2}d\phi\,
                                   \left[\Delta T_b(\theta,\phi)-
				         \Delta T_b(\theta,\phi+\pi/2)+
				   \right.\nonumber\\
				& & \;\;\;\;\;\;\;\;\;\;\;\;\;\;\;\;
				    \;\;\;\;\;\;\;\;\;\;\;\;\;\;\,
				    \left. \Delta T_b(\theta,\phi+\pi)-
				           \Delta T_b(\theta,\phi+3/2\pi)
                                           \right]\cdot f(\theta,\phi)\,,\\
f(\theta,\phi) &=& -P(\theta, \phi)\chi^*(\theta, \phi+\pi/2) +
                  \chi(\theta, \phi) P^*(\theta, \phi+\pi/2)\,,
\end{eqnarray}
where $P$ and $\chi$ are the co-polar and cross-polar patterns, respectively, 
and $\Omega_A$ is the antenna beam. 
In the case of SPOrt feed horns, the contribution of the $f$ pattern is 
$\sim -24$~dB and the {rms} contamination from the 30~$\mu$K of the 
CMB anisotropy  is lower than 0.2~$\mu$K.
Due to its intrinsic asimmetry, off-axis optics with the same cross-polar 
pattern level would imply a spurious contribution 8-10~dB higher.

Moreover, also the hardware calibration of a CMBP experiment represents a 
challenge. Standard
marker injectors are not suitable for calibrating a tensorial quantity as the
pair
($Q$,~$U$). Thus, a new concept calibrator has been developed,
valid for any radio-polarimeter,
based on the insertion of three signals at different position angles. This
device is similar to that of BaR-SPOrt and 
 further details can be found in Ref.~\citenum{Ba02,Za02}.

In summary, the faint CMBP signal requires specifically devoted 
instruments and the SPOrt team
has spent (and is still spending) a big effort in designing an 
instrument with very low
systematic error contamination, characterized by:
\begin{itemize}  
   \item correlation unit with high rejection of the unpolarized component 
         ($> 60$~dB) based on a custom-developed HPD and a lock-in system;
   \item on axis and simple optics (corrugated feed horns) in order to 
         minimize the spurious polarization
         induced by both the $f$ pattern 
         and the CMB temperature anisotropy at the beam scale: 
	 With such a configuration $\sim -35$~dB
         of cross-polarization translates into a contamination 
	 $<0.2$~$\mu$K;
   \item high OMT isolation ($> 60$~dB) and low cross-talk 
         ($< -60$~dB), since these parameters are 
         among the major responsibles for
         $Q$~\&~$U$ offset generation in correlation 
	 polarimeters;
   \item very small difference ($< -30$~dB) between the attenuations 
         of the two polarizations in the polarizer, which is 
         the other main responsible of offset generation. 
\end{itemize}

\section{Scientific Goals}

The goals of SPOrt are essentially two:
\begin{itemize}
   \item to provide $7^\circ$ (FWHM) full maps of the Galactic synchrotron 
         emission at 22-32~GHz.
The diffuse Galactic polarization, at frequencies greater than 2.7~GHz, 
is practically unknown. However, a level of
$T_{\rm syn}(30\,\,{\rm GHz})\sim 5\,\mu$K on 7$^{\circ}$ scales 
can be evaluated by down extrapolating data from Duncan et al.\cite{Du97}. 
Unpolarized data may provide an independent confirmation assuming, 
for example, a 30\% of polarization in the COBE-DMR maps.

\begin{figure} [htp]
\begin{center}
\includegraphics[width=0.55\linewidth, angle=0]{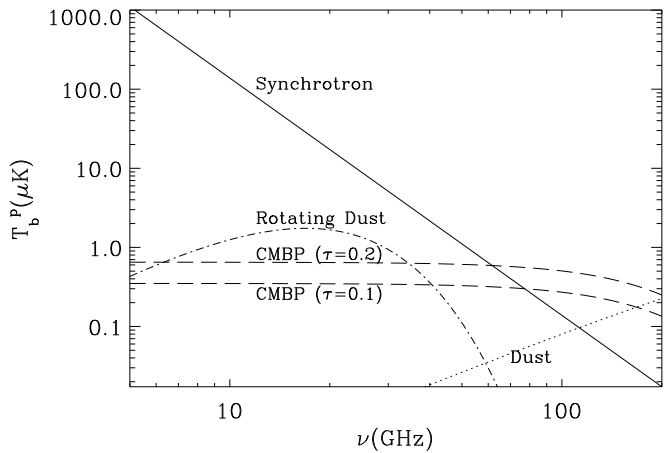}
\end{center}
\caption{Expected polarized brightness temperature for the relevant polarized
Galactic foregrounds on $7^{\circ}$ scale.
The synchrotron emission has been normalized to 
$T_{\rm syn}(30\,\,{\rm GHz})\sim 5\,\mu$K: See text for details. 
The
parameters of the other components are from Tegmark et al.\cite{Teg00}.
The
CMBP behaviour for two $\Lambda$CDM models ($\Omega_{\Lambda} = 0.7$) 
with optical depth $\tau = 0.1$ and  $\tau = 0.2$ are also shown.}
\label{galEmFig}
\end{figure}
The expected scenario is sketched in Figure~\ref{galEmFig},
where the CMBP is plotted 
together with the relevant foregrounds. 
The SPOrt sensitivities 
reported in Table~\ref{tabfeat} confirm that full maps of the 
Galaxy should be 
done at 22 and 32~GHz, following predictions. 

   \item to attempt a first detection of CMBP on large angular scales; 
         upper limits, at least 1 order of magnitude lower 
         than at present, can be achieved.
Since the SPOrt pixel sensitivities do not envisage the 
possibility constructing CMBP maps, only full-sky statistical 
analyses may provide an estimate of the mean polarized 
signal       
          $P_{\rm rms} = \sqrt{\left<Q^2 + U^2\right>}$.
Similarly to what has been done by PIQUE\cite{hedman02} and 
POLAR\cite{keating01} experiments by applying the flat spectrum 
analysis\cite{Zalda98}, simulations have shown that the CMBP measurement  
will have an error $\sigma(P_{\rm rms}) = 0.3$~$\mu$K (1$\sigma$ C.L.), 
taking into account the degradation from 
foreground subtraction.

Figure~4 of Ref.~\citenum{Ca02} shows the $P_{\rm rms}$ 
behaviour, with respect to $\tau$, is almost independent 
of other cosmological parameters. That is, the detection of the 
$P_{\rm rms}$ on large angular scales is relevant for a clean measurement 
of the optical depth of the re-ionized medium in the dark ages. A 
sensitivity of $\Delta\tau = 0.13$ for a model with $\tau=0.2$ 
has been determined 
through a Fisher matrix analysis.
\end{itemize}

\acknowledgments     
We thank V. Natale for useful discussions. 
We acknowledge use of CMBFAST package for 
performing our analysis. SPOrt is an ASI funded project.
We thank also ESA for provìding the flight opportunity onboard the 
International Space Station.
V.A.R., E.N.V., M.V.S. and I.A.S. are greatful to C.N.R. for partially
supporting their collaboration within the C.N.R.-R.A.S. agreement.


\end{document}